# PROPORTIONATE AFFINE PROJECTION ALGORITHMS FOR BLOCK-SPARSE SYSTEM IDENTIFICATION

*Jianming Liu and Steven L. Grant*[1]

Department of Electrical and Computer Engineering, Missouri University of Science and Technology, Rolla, Missouri 65409.

## ABSTRACT

A new family of block-sparse proportionate affine projection algorithms (BS-PAPA) is proposed to improve the performance for block-sparse systems. This is motivated by the recent block-sparse proportionate normalized least mean square (BS-PNLMS) algorithm. It is demonstrated that the affine projection algorithm (APA), proportionate APA (PAPA), BS-PNLMS and PNLMS are all special cases of the proposed BS-PAPA algorithm. Meanwhile, an efficient implementation of the proposed BS-PAPA and block-sparse memory PAPA (BS-MPAPA) are also presented to reduce computational complexity. Simulation results demonstrate that the proposed BS-PAPA and BS-MPAPA algorithms outperform the APA, PAPA and MPAPA algorithms for block-sparse system identification in terms of both faster convergence speed and better tracking ability.

*Index Terms* —Proportionate affine projection algorithm, block-sparse system identification, adaptive filter

## I. INTRODUCTION

The impulse responses of many applications, such as network echo cancellation (NEC), are sparse, which means a small percentage of the impulse response components have a significant magnitude while the rest are zero or small. Therefore, instead of the normalized least mean square (NLMS) [1] and the affine projection algorithm (APA) [2], the family of proportionate algorithms exploits this sparseness to improve their performance, including proportionate NLMS (PNLMS) [3], and proportionate APA (PAPA) [4]. The memory improved PAPA (MIPAPA) algorithm was proposed to not only speed up the convergence rate but also reduce the computational complexity by taking into account the memory of the proportionate coefficients [5].

It has been shown that the PNLMS algorithm and PAPA can both be deduced from a basis pursuit perspective [6]-[7]. A more general framework was further proposed to derive the PNLMS adaptive algorithms for sparse system identification, which employed convex optimization [8]. Recently, the block-sparse PNLMS (BS-PNLMS) algorithm was proposed to improve the performance of PNLMS for identifying block-sparse systems [9]. Motivated by BS-PNLMS, we propose a family of block-sparse PAPA algorithms for block-sparse system identification in this paper. The PNLMS, BS-PNLMS, APA and PAPA algorithms are all special cases of this proposed BS-PAPA algorithm. Meanwhile, in order to reduce the computational complexity, taking advantage of the block-sparse property in the proposed BS-PAPA algorithm, an efficient implementation of BS-PAPA is studied, and the block-sparse memory PAPA (BS-MPAPA) algorithm is also introduced.

## II. REVIEW OF PAPA

In the typical echo cancellation problem, the input signal $\mathbf{x}(n)$ is filtered through the unknown coefficients $\mathbf{h}(n)$ to get the observed output signal $d(n)$.

$$d(n) = \mathbf{x}^T(n)\mathbf{h}(n) + v(n), \qquad (1)$$

where

$$\mathbf{x}(n) = [x(n), x(n-1), \cdots, x(n-L+1)]^T,$$

$v(n)$ is the measurement noise, and $L$ is the length of impulse response. We define the estimated error as

$$e(n) = d(n) - \mathbf{x}^T(n)\hat{\mathbf{h}}(n-1), \qquad (2)$$

where $\hat{\mathbf{h}}(n)$ is the adaptive filter's coefficients. Grouping the $M$ most recent input vectors $\mathbf{x}(n)$ together gives the input signal matrix:

$$\mathbf{X}(n) = [\mathbf{x}(n), \mathbf{x}(n-1), \cdots, \mathbf{x}(n-M+1)]$$

Therefore, the estimated error vector is

$$\mathbf{e}(n) = \mathbf{d}(n) - \mathbf{X}^T(n)\hat{\mathbf{h}}(n-1), \qquad (3)$$

in which

---

$$\mathbf{d}(n) = [d(n)\, d(n-1) \cdots d(n-M+1)],$$

$$\mathbf{e}(n) = [e(n)\, e(n-1) \cdots e(n-M+1)],$$

and $M$ is the projection order. The PAPA algorithm updates the filter coefficients as follows [4]:

$$\hat{\mathbf{h}}(n) = \hat{\mathbf{h}}(n-1) + \mu \mathbf{G}(n-1)\mathbf{X}(n)\left(\mathbf{X}(n)\mathbf{G}(n-1)\mathbf{X}(n) + \delta \mathbf{I}_M\right)^{-1} \mathbf{e}(n), \quad (4)$$

in which $\mu$ is the step-size, $\delta$ is the regularization, $\mathbf{I}_M$ is $M \times M$ identity matrix and

$$\mathbf{G}(n-1) = diag\left[g_1(n-1), g_2(n-1), \cdots, g_L(n-1)\right], \quad (5)$$

$$g_l(n-1) = \frac{\gamma_l(n-1)}{\frac{1}{L}\sum_{i=1}^{L}\gamma_i(n-1)}, \quad (6)$$

$$\gamma_l = \max\left\{\rho \max\left\{q, |\hat{h}_1|, \cdots, |\hat{h}_L|\right\}, |\hat{h}_l|\right\}, \quad (7)$$

$q$ prevents the filter coefficients $\hat{h}_l(n-1)$ from stalling when $\hat{\mathbf{h}}(0) = \mathbf{0}_{L \times 1}$ at initialization, and $\rho$ prevents the coefficients from stalling when they are much smaller than the largest coefficient.

In many applications, including network echo cancellation (NEC) and satellite-linked communication echo cancellation, the impulse response is block sparse, that is, it consists of several dispersive active regions. However, PAPA does not take into account the block-sparse characteristic, and motivated by the block-sparse PNLMS (BS-PNLMS) algorithm [9], we propose a family of new block-sparse PAPA algorithms to further improve their performance for identifying the block-sparse impulse system in next section.

### III. PROPOSED BS-PAPA

The block-sparse scheme for PAPA will be firstly derived based on the optimization of $l_{2,1}$ norm, then in order to reduce the computational complexity, an efficient implementation of the proposed BS-PAPA is presented by taking advantage of the block structure. Finally, block-sparse memory PAPA (BS-MPAPA) is also proposed by considering the memory of the coefficients to further reduce computational complexity.

#### A. The Proposed BS-PAPA

The proportionate APA algorithm can be deduced from a *basis pursuit* perspective as below [7]

$$\begin{aligned}&\textbf{\textit{min}} \quad \|\tilde{\mathbf{h}}(n)\|_1 \\ &\text{subject to} \quad \mathbf{d}(n) = \mathbf{X}^T(n)\tilde{\mathbf{h}}(n),\end{aligned} \quad (8)$$

where $\tilde{\mathbf{h}}(n)$ is the correction component defined as [6]-[7]

$$\tilde{\mathbf{h}}(n) = \mathbf{G}(n-1)\mathbf{X}(n)\left[\mathbf{X}^T(n)\mathbf{G}(n-1)\mathbf{X}(n)\right]^{-1}\mathbf{d}(n). \quad (9)$$

Motivated by BS-PNLMS, the proposed block-sparse scheme for PAPA is derived by replacing the $l_1$ norm optimization target in the *basis pursuit* perspective with the following $l_{2,1}$ norm defined as

$$\|\tilde{\mathbf{h}}\|_{2,1} = \left\|\begin{array}{c}\|\tilde{\mathbf{h}}_{[1]}\|_2 \\ \|\tilde{\mathbf{h}}_{[2]}\|_2 \\ \vdots \\ \|\tilde{\mathbf{h}}_{[N]}\|_2\end{array}\right\|_1 = \sum_{i=1}^{N}\|\tilde{\mathbf{h}}_{[i]}\|_2, \quad (10)$$

where $\|\tilde{\mathbf{h}}_{[i]}\|_2 = \sqrt{\tilde{\mathbf{h}}_{[i]}^T \tilde{\mathbf{h}}_{[i]}}$, $\tilde{\mathbf{h}}_{[i]} = [\tilde{h}_{(i-1)P+1}, \tilde{h}_{(i-1)P+2}, \cdots, \tilde{h}_{iP}]^T$, $P$ is a predefined group partition size parameter and $N = L/P$ is the number of groups. Therefore,

$$\begin{aligned}&\textbf{\textit{min}} \quad \|\tilde{\mathbf{h}}(n)\|_{2,1} \\ &\text{subject to} \quad \mathbf{d}(n) = \mathbf{X}^T(n)\tilde{\mathbf{h}}(n).\end{aligned} \quad (11)$$

Similarly, the proposed BS-PAPA could be derived using the method of Lagrange multipliers, see [6]-[7] for more details. The update equation for the proposed BS-PAPA is then,

$$\hat{\mathbf{h}}(n) = \hat{\mathbf{h}}(n-1) + \mu \mathbf{G}(n-1)\mathbf{X}(n)\left(\mathbf{X}^T(n)\mathbf{G}(n-1)\mathbf{X}(n) + \delta \mathbf{I}_M\right)^{-1}\mathbf{e}(n), \quad (12)$$

and

$$\mathbf{G}(n-1) = diag\left[\|\hat{\mathbf{h}}_{[1]}\|_2 \mathbf{1}_P, \|\hat{\mathbf{h}}_{[2]}\|_2 \mathbf{1}_P, \cdots, \|\hat{\mathbf{h}}_{[N]}\|_2 \mathbf{1}_P\right], \quad (13)$$

in which $\mathbf{1}_P$ is a $P$-length row vector of all ones. Equation (12) is the same as traditional PAPA in (4), except for the block-sparse definition of $\mathbf{G}(n-1)$ in (13). Similar to (5)-(7) in PAPA to prevent the stalling issues, the proposed BS-PAPA replaces (5)-(7) with

$$\mathbf{G}(n-1) = diag\left[g_1(n-1)\mathbf{1}_P, g_2(n-1)\mathbf{1}_P, \cdots, g_N(n-1)\mathbf{1}_P\right], \quad (14)$$

$$g_i(n-1) = \frac{\gamma_i}{\frac{1}{N}\sum_{l=1}^{N}\gamma_l}, \quad (15)$$

$$\gamma_i = \max\left\{\rho \max\left\{q, \|\hat{\mathbf{h}}_{[1]}\|_2, \cdots, \|\hat{\mathbf{h}}_{[N]}\|_2\right\}, \|\hat{\mathbf{h}}_{[i]}\|_2\right\}. \quad (16)$$

It should be noted that the proposed BS-PAPA includes PNLMS, BS-PNLMS, APA and PAPA. The BS-PNLMS algorithm is a special case of BS-PAPA with projection order $M = 1$. In the case of $P$ is equal to 1, the BS-PAPA algorithm degenerates to PAPA. Meanwhile, when $P$ is chosen as $L$, the proposed BS-PAPA turns into APA.

*B. Efficient implementation of proposed BS-PAPA*

By taking advantage of the new block-sparse characteristic in the proposed BS-PAPA algorithm, we can reduce the computational complexity of the proposed BS-PAPA, especially for higher projection order. Equation (12) can be rewritten as

$$\mathbf{P}(n) = \mathbf{G}(n-1)\mathbf{X}(n), \quad (17)$$

$$\hat{\mathbf{h}}(n) = \hat{\mathbf{h}}(n-1) + \mu \mathbf{P}(n)\left(\mathbf{X}^T(n)\mathbf{P}(n) + \delta \mathbf{I}_M\right)^{-1} \mathbf{e}(n). \quad (18)$$

Considering the blocks of $\mathbf{G}(n-1)$ in (14), (17) can be rewritten as (19) below, where

$$\mathbf{x}_P(n) = [x(n)\, x(n-1) \cdots x(n-P+1)]^T \quad (20)$$

The direct implementation of (17) will need $ML$ multiplications, which is the case of classical PAPA. However, considering the block-sparse characteristic in (14), the computational complexity of (19) can be further reduced. The $i$th submatrix of $\mathbf{P}(n)$ is defined as $\mathbf{P}_i(n)$ in (21) at the bottom of the page. Considering the shift property of $\mathbf{x}_P(n)$ in (20), we only need to calculate the vector $\mathbf{p}_i(n)$ in (22) at the bottom of the page which requires $P+M-1$ multiplications then use a sliding window to construct $\mathbf{P}_i(n)$. Therefore, the number of multiplications of (19) in the proposed BS-PAPA will become $(P+M-1)N$.

It should be noted that, the proposed efficient implementation will not damage the performance of the BS-PAPA algorithm. Meanwhile, the advantage of proposed efficient implementation becomes more apparent when the projection order and block size increase.

*C. Memory BS-PAPA*

In order to further reduce the computational complexity of (19), we could consider the memory of proportionate coefficients as in [5], and approximate the matrix $\mathbf{P}(n)$ by $\mathbf{P}'(n)$ in (23) at the bottom of the page. Due to time-shift property of (23), it could be implemented as

$$\mathbf{P}'(n) = \left[\mathbf{g}(n-1) \odot \mathbf{x}(n), \mathbf{P}'_{-1}(n-1)\right] \quad (24)$$

where the operation $\odot$ denotes the Hadamard product and the matrix $\mathbf{P}'_{-1}(n-1)$ contains the first $M-1$ columns of $\mathbf{P}'(n-1)$. The calculation of $\mathbf{P}'(n)$ only needs $L$ multiplications and the proposed BS-MPAPA updates the coefficients as below:

$$\hat{\mathbf{h}}(n) = \hat{\mathbf{h}}(n-1) + \mu \mathbf{P}'(n)\left(\mathbf{X}^T(n)\mathbf{P}'(n) + \delta \mathbf{I}_M\right)^{-1} \mathbf{e}(n). \quad (25)$$

It should be noted that the efficient implementation proposed in Section III.B could not be applied to the memory BS-PAPA, however, the computational complexity of memory BS-PAPA will be lower than BS-PAPA due to the time-shift property when considering the memory.

$$\mathbf{P}(n) = \begin{bmatrix} g_1(n-1)\mathbf{x}_P(n), & g_1(n-1)\mathbf{x}_P(n-1), & \cdots, & g_1(n-1)\mathbf{x}_P(n-M+1) \\ g_2(n-1)\mathbf{x}_P(n-P), & g_2(n-1)\mathbf{x}_P(n-P-1), & \cdots, & g_2(n-1)\mathbf{x}_P(n-M+1-P) \\ \vdots & \vdots & \ddots & \vdots \\ g_N(n-1)\mathbf{x}_P(n-(N-1)P), & g_N(n-1)\mathbf{x}_P(n-(N-1)P-1), & \cdots, & g_N(n-1)\mathbf{x}_P(n-M+1-(N-1)P) \end{bmatrix} \quad (19)$$

$$\mathbf{P}_i(n) = \left[g_i(n-1)\mathbf{x}_P(n-(i-1)P),\; g_i(n-1)\mathbf{x}_P(n-(i-1)P-1),\; \cdots,\; g_i(n-1)\mathbf{x}_P(n-M+1-(i-1)P)\right] \quad (21)$$

$$\mathbf{p}_i(n) = \left[g_i(n-1)x(n-(i-1)P),\; g_i(n-1)x(n-(i-1)P-1),\; \cdots,\; g_i(n-1)x(n-M+2-iP)\right]^T \quad (22)$$

$$\mathbf{P}'(n) = \begin{bmatrix} g_1(n-1)\mathbf{x}_P(n), & g_1(n-2)\mathbf{x}_P(n-1), & \cdots, & g_1(n-M)\mathbf{x}_P(n-M+1) \\ g_2(n-1)\mathbf{x}_P(n-P), & g_2(n-2)\mathbf{x}_P(n-P-1), & \cdots, & g_2(n-M)\mathbf{x}_P(n-M+1-P) \\ \vdots & \vdots & \ddots & \vdots \\ g_N(n-1)\mathbf{x}_P(n-(N-1)P), & g_N(n-2)\mathbf{x}_P(n-(N-1)P-1), & \cdots, & g_N(n-M)\mathbf{x}_P(n-M+1-(N-1)P) \end{bmatrix} \quad (23)$$

## IV. SIMULATION RESULTS

The performance of the proposed BS-PAPA and BS-MPAPA are evaluated via simulations. Throughout our simulation, the length of the unknown system is $L = 1024$, and the adaptive filter is the same length. Two block-sparse impulse systems in Figure 1 are used: the first impulse response in Figure 1(a) is with a single cluster of nonzero coefficients at [257, 288], which has 32 taps; the two clusters in the second impulse response in Figure 1(b) locate at [257, 288] (32 taps) and [769, 800] (32 taps) separately. In order to compare the tracking ability for different algorithms, an echo path change was incurred at 30000-sample by switching from the first impulse response in Figure 1(a) to the second impulse response in Figure 1(b).

The algorithms were tested using colored noise which was generated by filtering white Gaussian noise (WGN) through a first order system with a pole at 0.8. Independent WGN is added to the system background with a signal-to-noise ratio, SNR = 30dB. The projection order was $M = 8$, and the step-sizes were $\mu = 0.01$. The regularization parameters $\delta$ were set to 0.01, and we used $\rho = 0.01$, and $q = 0.01$. The convergence state of adaptive filter is evaluated with the normalized misalignment which is defined as $10\log_{10}(\|\mathbf{h} - \hat{\mathbf{h}}\|_2^2 / \|\mathbf{h}\|_2^2)$.

The performance of the proposed BS-PAPA was tested for different group sizes $P$ chosen as 1 (i.e. PAPA), 4, 16, 32, 64, 1024 (i.e. APA) separately in Figure 2. The impact of different group sizes on BS-MPAPA is similar. As discussed in BS-PNLMS [9], the group size should be chosen properly (around 32 here) in order to fully take advantage of the block-sparse characteristic.

In the second simulation, we compare the performance of BS-PAPA and BS-MPAPA algorithms together with APA, PAPA and MPAPA. For both the BS-PAPA and BS-MPAPA algorithms, the group size was $P = 32$. The convergence curves for colored input are shown in Figure 3. As can be seen, both proposed BS-PAPA and BS-MPAPA outperform PAPA and MPAPA in terms of convergence speed and tracking ability. Meanwhile, BS-MPAPA will be more favorable considering its lower computational complexity.

## V. CONCLUSION

We have proposed two proportionate affine projection algorithms for block-sparse system identification, called block-sparse PAPA (BS-PAPA) and block-sparse memory PAPA (BS-MPAPA). Simulation results demonstrate that the new proportionate BS-PAPA and BS-MPAPA algorithms outperform traditional PAPA, MPAPA for block-sparse system identification.

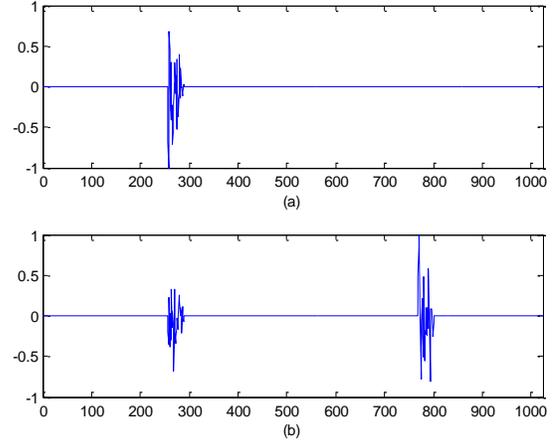

**Figure 1** Block-sparse impulse systems (a) one-cluster block-sparse system, (b) two-cluster block-sparse system.

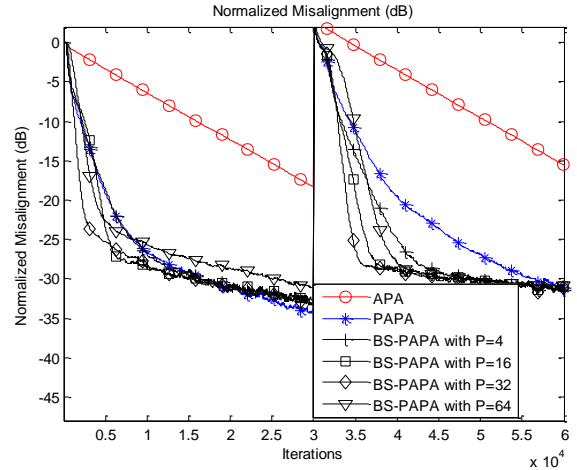

**Figure 2** Comparison of BS-PAPA with different group sizes for colored input with $SNR$=30dB.

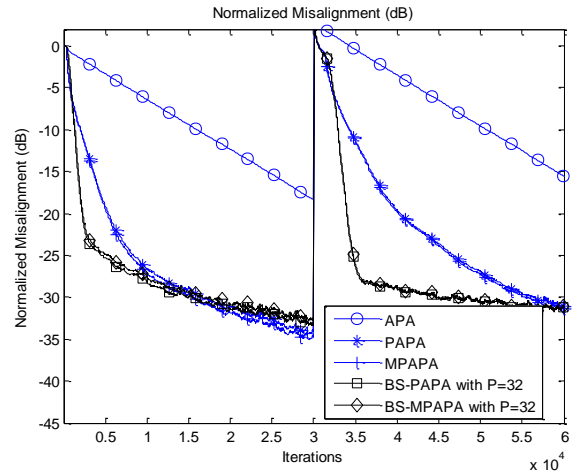

**Figure 3** Comparison of APA, PAPA, MPAPA, BS-PAPA and BS-MPAPA algorithms for colored noise with $SNR$=30dB.